\documentclass[
    ,final            
  ]
  {aipproc}

\layoutstyle{6x9}

\newcommand{\Dphi}{$\Delta \phi$}
\newcommand{\ptmax}{$p_T^{{\rm max}}$}
\newcommand{\pt}{$p_T$}

\begin{document}

\title{D0 Measurement of the Dijet Azimuthal Decorrelations}


\author{Marek Zieli\'nski\footnote{For the D0 Collaboration.}}{
  address={University of Rochester, Rochester, NY, U.S.A.}
}

%

\begin{abstract}

We present the D0 measurement of correlations in the azimuthal 
angle between the two largest
transverse momentum jets produced
in $p\bar{p}$ collisions at a center-of-mass energy
$\sqrt{s}=1.96$ TeV in the
central rapidity region \cite{dphi}.
The results are based on an inclusive dijet event sample
corresponding to an integrated luminosity of
150 pb$^{-1}$. Data is in good agreement with next-to-leading
order (NLO) pQCD calculations, and with tuned {\sc pythia, herwig, alpgen}
and {\sc sherpa} event generators.

\end{abstract}

\maketitle


The proper description of multi-parton radiation
is crucial for a wide range of precision measurements as
well as for searches for new physical phenomena.
The azimuthal difference \Dphi\ between two leading jets 
in an event is a clean and simple probe of radiation effects.
In the absence of radiation \Dphi = $\pi$; soft radiation causes
small deviations from $\pi$, while \Dphi\ significantly lower
than $\pi$ indicates the presence of hard radiation, such as
a third jet with high \pt. The range \Dphi < $2\pi/3$ is populated
by events with 4 or more hard jets
Distributions in \Dphi\ allow testing the QCD descriptions
across a range of jet multiplicities
without requiring the reconstruction
of additional jets
and offer a way to examine the
transition between soft and hard QCD processes based on a
single observable.
The observable is defined as the differential dijet cross section
in \Dphi, normalized by the dijet cross section integrated over
\Dphi\ in the same phase space
$(1/\sigma_{\rm dijet})\, (d\sigma_{\rm dijet}/ d\Delta\phi)$.
Theoretical and experimental uncertainties are reduced in this
construction.
Jets
are defined using an iterative seed-based cone algorithm (including
mid-points) with radius $R_{\rm cone}=0.7$~\cite{run2cone} at
parton, particle, and experimental levels.

Four analysis regions were defined based on the jet with largest
$p_T$ in an event (\ptmax) with the requirement that the trigger
efficiency be at least 99\%.
The second leading $p_T$ jet in each event was required to have
$p_T>40$~GeV and both jets were required to have central rapidities with
$|y| < 0.5$.
Several event selection criteria reduced effects of misreconstruction
and the backgrounds from cosmic rays, as well as from electrons,
photons and noise mimicking jets \cite{dphi}.
The overall selection efficiency was typically $\approx 80$\%.
The jet energy calibration was the largest contribution to
systematic uncertainty.

\begin{figure}
\hbox{
  \includegraphics[width=.5\textwidth]{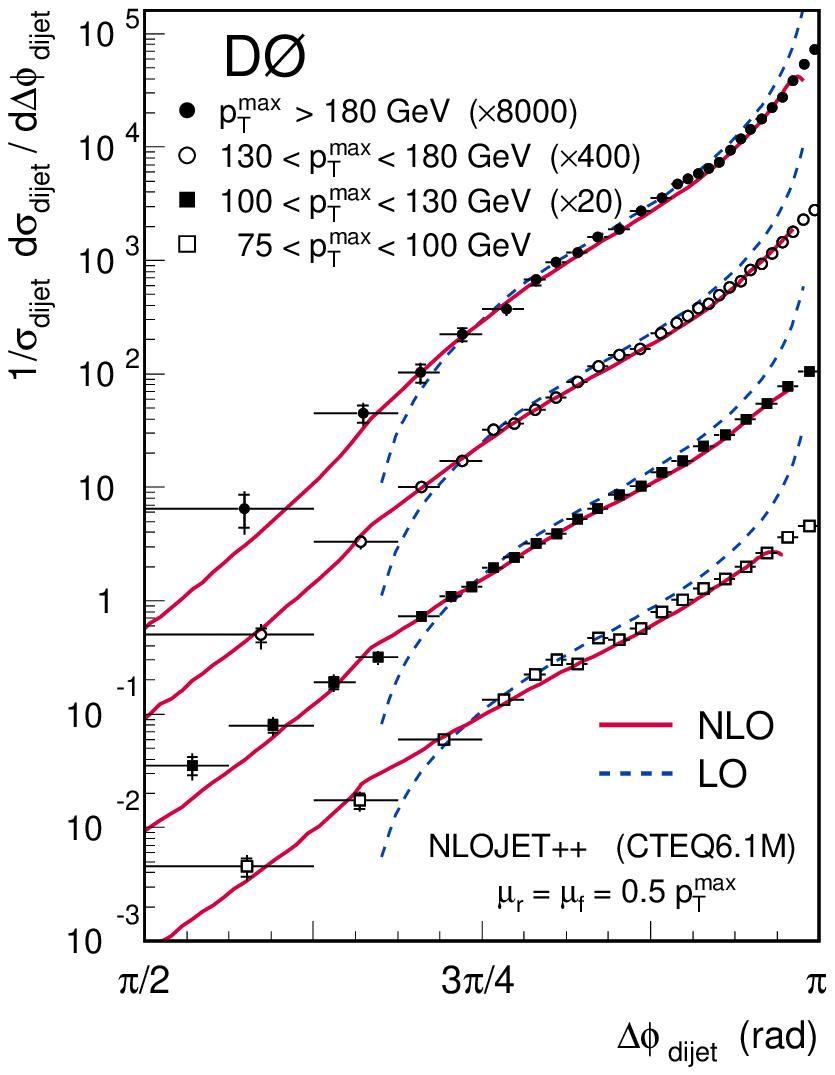}
  \includegraphics[width=.5\textwidth]{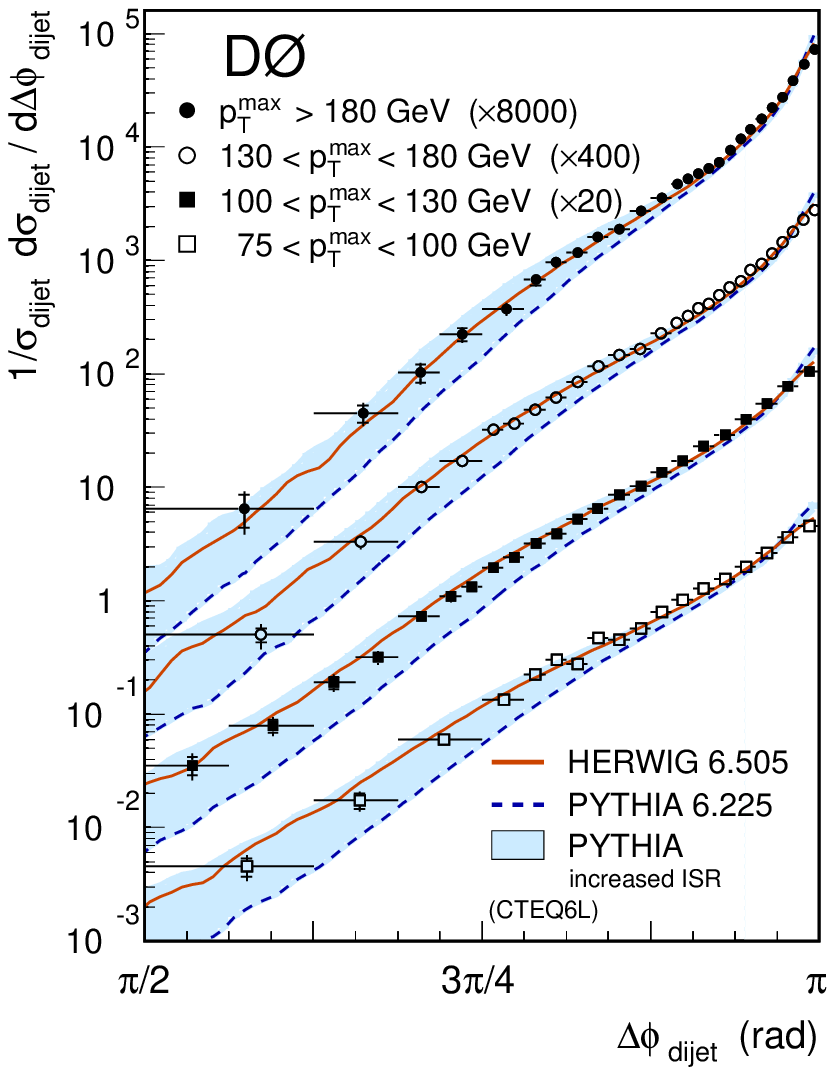}
}
  \caption{
Left: The \Dphi\ distributions in different \ptmax\ ranges.
Data and predictions with \ptmax\ > 100 GeV are scaled by
successive factors of 20 for purposes of presentation.
The solid (dashed) lines show the NLO (LO) pQCD predictions.
Right: Analogous presentation of results from {\sc herwig} and 
{\sc pythia} overlaid on the data.
}
\label{fig:NLOandMC}
\end{figure}

The corrected data are presented in Fig.~\ref{fig:NLOandMC} as a function
of \Dphi\ in four ranges of \ptmax.
The spectra are strongly peaked at \Dphi$\approx\pi$;
the peaks are narrower at larger values of \ptmax.
Overlaid on the data points in Fig.~\ref{fig:NLOandMC} (left) 
are the results of pQCD
calculations obtained using the parton-level event generator {\sc
nlojet++}~\cite{nlojet};
NLO pQCD provides a good description of data, while the leading-order
(LO) calculation has a limited applicability.
The pQCD calculations are insensitive to hadronization corrections 
and the underlying event effects \cite{wobisch}.

Monte Carlo event generators, such as {\sc herwig}~\cite{herwig} 
and {\sc pythia}~\cite{pythia},
use $2 \rightarrow 2$  LO pQCD matrix elements with phenomenological
parton-shower models to simulate higher order QCD effects.
Results from {\sc herwig} (version 6.505) and {\sc pythia} (version 6.225),
both using default parameters,
are compared
to the data in Fig.~\ref{fig:NLOandMC} (right).
{\sc herwig} 
describes the data well over the
entire \Dphi\ range including \Dphi$\approx\pi$.
{\sc pythia} with default parameters describes the data poorly---the
distribution is too narrowly peaked at \Dphi$\approx\pi$ and lies
significantly below the data over most of the \Dphi\ range.
The shaded bands in Fig.~\ref{fig:NLOandMC} (right) indicate 
the range of variation
when the maximum allowed virtuality in the initial-state shower
is smoothly increased from the
current default by a factor of four.
These variations 
clearly demonstrate the sensitivity of this measurement.
Global efforts to tune Monte Carlo event generators
can benefit from including our data to constrain the 
related model parameters.

\begin{figure}
\hbox{
  \includegraphics[width=.5\textwidth]{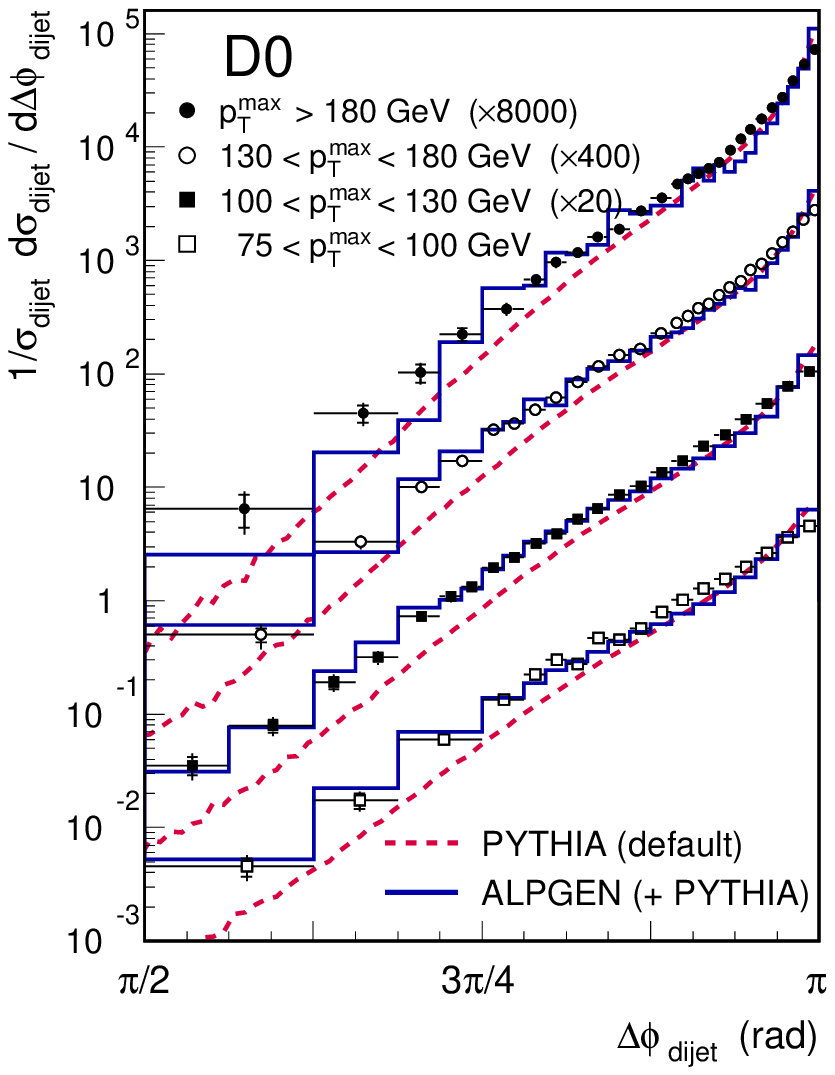}
  \includegraphics[width=.5\textwidth]{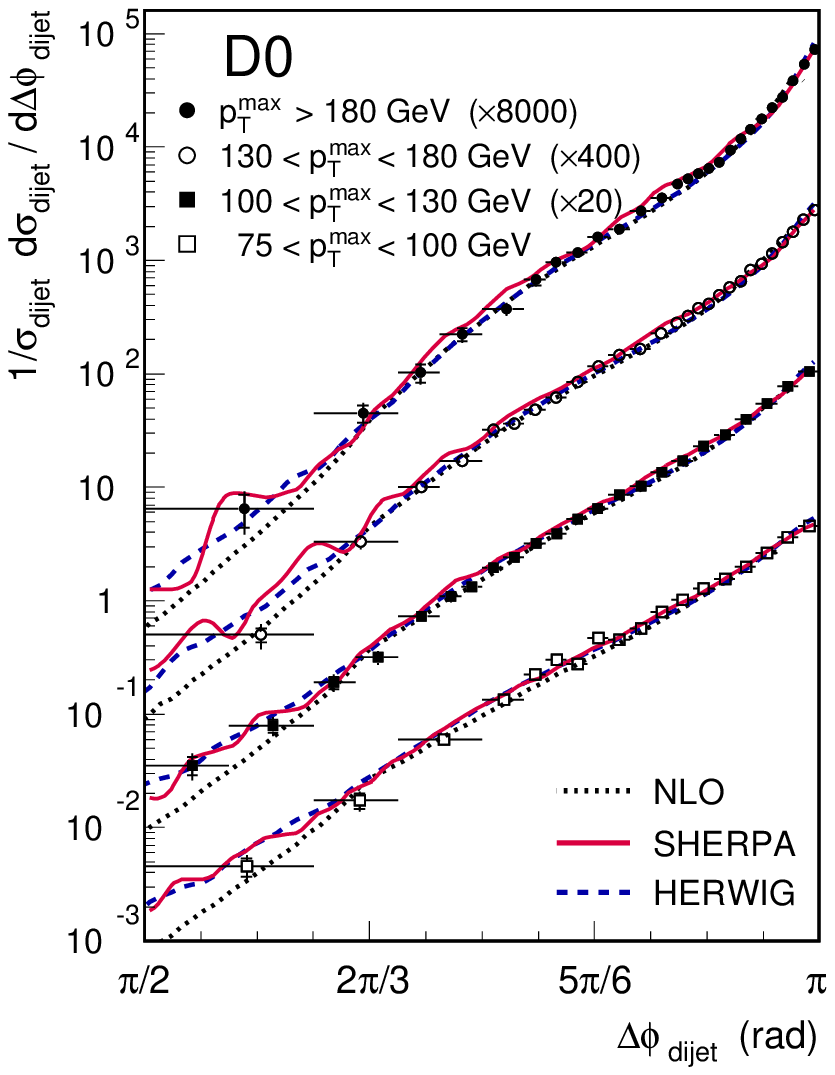}
}
  \caption{
Left: Comparison of predictions from {\sc alpgen} (interfaced
to default {\sc pythia}) with data.
Right: Comparison of predictions from {\sc sherpa} to data,
and to predictions from {\sc nlojet++} and from {\sc herwig}. 
}
\label{fig:AnS}
\end{figure}

Recent developments in event generators aim at improving the description
of processes involving multiple jets. These approaches combine
exact LO PQCD matrix elements for multi-parton production with
parton-shower models; care must be taken to avoid double counting
of equivalent phase-space configurations. The \Dphi\ distributions
are sensitive to a range of jet multiplicities and provide a test of 
such techniques. Two such generators, {\sc alpgen}~\cite{alpgen} 
and {\sc sherpa}~\cite{sherpa},
are shown to be in good agreement with data (Fig.~\ref{fig:AnS}),
thus enhancing confidence in their applications to other processes.



\begin{theacknowledgments}

We thank W. Giele, T. Gleisberg, F. Krauss, M. Mangano, Z. Nagy, M.~H. Seymour,
and T. Sj\"ostrand for discussions.

\end{theacknowledgments}



\bibliographystyle{aipproc}   

\bibliography{sample}





\end{document}